\documentclass[10pt]{article}
\usepackage{amssymb,amscd,amsmath,amsthm}
\usepackage{latexsym,amstext}
\usepackage{latexsym,amstext}
\usepackage{color}
\usepackage{graphicx}
\usepackage{epstopdf}
\usepackage{lineno,hyperref}
\modulolinenumbers[5]

\makeatletter
\@addtoreset{equation}{section}
\makeatother

\begin{document}

\title{Two-point one-dimensional $\delta$-$\delta^\prime$ interactions: \\ \vspace{0.15cm} 
non-abelian addition law and decoupling limit}

 \author{M. Gadella$^1$, J. Mateos-Guilarte$^2$, J.M. Mu\~noz-Casta\~neda$^3$\thanks{Corresponding author; jose.m.munoz.castaneda@gmail.com}, L.M. Nieto$^1$}

\maketitle

\noindent
$^1$ Departamento de F\'{\i}sica Te\'orica, At\'omica y
\'Optica, and IMUVA (Instituto de Matem\'aticas), 
Universidad de Va\-lladolid, 47011 Valladolid, Spain
\smallskip

\noindent
$^2$ Departamento de F\'isica Fundamental, Universidad de Salamanca, 37008 Salamanca, Spain
\smallskip

\noindent
$^3$ Institut f\"ur Theoretische Physik, Universit\"at Leipzig, Germany and Instituto de Ciencia de Materiales de Arag\'on-Departamento de F\'\i sica de la Materia Condensada, CSIC-Universidad de Zaragoza, E-50012 Zaragoza, Spain

\bigskip

\begin{abstract}
In this contribution to the study of one dimensional point potentials, we prove that if we take the limit $q\to 0$ on a potential of the type $v_0\delta({y})+{ 2}v_1\delta'({y})+w_0\delta({ y}-q)+ { 2} w_1\delta'({y}-q)$, we obtain a new point potential of the type ${ u_0} \delta({ y})+{ 2 u_1} \delta'({ y})$, when $ u_0$ and $ u_1$ are related to $v_0$, $v_1$, $w_0$ and $w_1$ by a law having the structure of a group. This is the Borel subgroup of $SL_2({\mathbb R})$. We also obtain the non-abelian addition law from the scattering data. The spectra of the Hamiltonian in the decoupling cases emerging in the study are also described in full detail. It is shown that for the $v_1=\pm 1$, $w_1=\pm 1$ values of the $\delta^\prime$ couplings the singular Kurasov matrices become equivalent to Dirichlet at one side of the point interaction and Robin boundary conditions at the other side.
\end{abstract}


\section{Introduction}

One dimensional models with point interactions \cite{AL} have recently received much attention. They serve to modeling several kinds of extra thin structures \cite{Z1,Z2} or point defects in materials, so that effects like tunneling are easily studied. They are also interesting in the study of heterostructures, where they may appear in connection to an abrupt effective mass change \cite{GHNN}. The general study of point interactions of the free Hamiltonian $H_0=-\frac{\hbar^2}{2m}\,\frac{d^2}{dx^2}$ is mainly due to Kurasov \cite{K,K1} and it is based on the construction of self adjoint extensions of symmetric operators with identical deficiency indices. More recently Asorey, Mu\~noz-Casta\~neda and coworkers reformulated the theory of self adjoint extensions of symmetric operators over bounded domains in terms of meaningful quantities from a quantum field theoretical point of view, see Refs. \cite{aso-mc1,aso-mc2,mc-kk1} and references therein. This new approach allows a rigorous study of the theory of quantum fields over bounded domains and the quantum boundary effects. Here, it is well known the existence of a four parameter family of self adjoint extensions. Some of these extensions are customarily associated to Hamiltonians of the type $H_0$ plus an interaction of type $a\delta(x-d)+b\delta'(x-d)$, where $\delta'$ is the derivative of the Dirac delta while $a$, $b$ and $d$ are fixed real numbers. 
This type of perturbation has interest both mathematical and physical in Quantum Mechanics and has been largely discussed \cite{AL,GOL,CAL,GNN,SE,FAS,Za1,Za2,Za3}. 

Although it is totally clear which self-adjoint extension corresponds to the 1D-Dirac delta $\delta(x-d)$, there is no consensus on which one should be assigned to its derivative $\delta'(x-d)$  \cite{SE,FAS,Za1,Za2,Za3,FAR,KP}. Self adjoint extensions of $H_0$ are characterized by matching conditions at $x=d$. The set of self adjoint extensions of $H_0$ is given by the unitary group $U(2)$ and therefore is a 4-parameter family of operators (see Refs. \cite{aso-mc2,mc-kk1}). In previous works, we have characterized perturbations of type $a\delta(x-d)+b\delta'(x-d)$ by suitable matching conditions as a two parameter family of self adjoint extensions (see Refs. \cite{GNN,MM}). Combining this family of point interactions with some other potentials (or even with mass jumps), we have obtained some physical features such that transmission and reflection coefficients, bound and antibound states and resonance poles \cite{AGLM}.

In other physical context, scalar QFT on a line, point potentials are useful to model impurities and/or providing external singular backgrounds where the bosons move, see e.g. \cite{JMG}. The spectra of Hamiltonians with $\delta$ and $\delta^\prime$ point interactions provide one-particle states in scalar $(1+1)$-dimensional QFT systems, see Refs \cite{aso-mc1,aso-mc2,mc-kk1}. In particular, configurations of two pure delta potentials added to the free Schr$\ddot{\rm o}$dinger Hamiltonian have found applications to describe scalar field fluctuations on external backgrounds, see e.g. \cite{JMC,bmc1,bor1}, as the corresponding scattering waves. The same configuration of delta interactions is addressed in Reference \cite{FAR} as a perturbation of the Salpeter Hamiltonian. Moreover, according to the idea proposed by several authors, delta point interactions allow to implement some boundary conditions compatible with an scalar QFT defined on an interval, see \cite{JMG} and References quoted therein.  Other than these generalized Dirichlet boundary conditions were discussed in \cite{MM} where it is shown that the use of $\delta$-$\delta^\prime$ potentials provides a much larger set of admissible boundary conditions. The $TGTG$-formula was subsequently applied to compute the corresponding quantum vacuum energies between two plane parallel plates represented by a $\delta$-$\delta^\prime$ potential in arbitrary space-time dimension.

To be more specific, the coupling $a$ to the $\delta$ potential mathematically describes the plasma frequencies in Barton's hydrodynamical model \cite{Barton2004} characterizing the electromagnetic properties of the conducting (infinitely thin) plates. The physical meaning of the $b$ coupling to the $\delta^\prime$ interaction in the context of Casimir physics has been unveiled only recently in \cite{bordag14-prd89}: it describes the response of the orthogonal polarizability of a monoatomically thin plate to the electromagnetic field.

Since the zero range potentials mimick the plates in a Casimir effect setup it is
interesting to consider a three-plate configuration and investigate the Casimir forces by allowing to move freely the plate in the middle, see e.g. \cite{Zee} for an introduction to Casimir Pistons, and \cite{fucci1,fucci2,fucci3,fucci4} for recent results. In particular, it is meaningful to displace the central plate towards one of the other two placed in the boundary; thus, we find the main physical motivation to study the particular situation where two $\delta$-$\delta^\prime$ interactions are superimposed.

This is the first problem to be analyzed in this work. The outcome is surprising: a non-abelian addition law emerges which corresponds to the Borel subgroup of the $SL_2({\mathbb R})$ group. The other focus of interest is the case where the $\delta^\prime$-couplings are exceptional, i.e., those couplings for which the transmission coefficients are zero and the plates become completely opaque: the left-right decoupling limit. It happens that the distinguished Dirichlet/Robin boundary conditions are implemented in this case by the $\delta$-$\delta^\prime$-potentials but the superposition law just mentioned becomes singular. We shall fully discuss this much more awkward regime in the second part of the paper.
 
Here, we shall combine two point potentials of the type $a_d\delta(x-d)+2b_d\delta'(x-d)$ with $d>0$, and  $a_0\delta(x)+2b_0\delta'(x)$. In fact, the idea of using double delta potentials ($b_d=b_0=0$) has a long tradition. For instance, in condensed matter physics, in the BCS model \cite{COO}, in Bose Einstein condensates \cite{BE}, or potentials with double pole resonances \cite{MON} arise in the description of some unstable states. Finally, regarding physical contexts where $\delta$-$\delta^\prime$ interactions play a r$\hat{\rm o}$le, we mention that point supported potentials can be used to model gap-like impurities in graphene layers and nano-ribbons, see Ref. \cite{lmm}. In this physical situations graphene surface plasmons suffer total reflection after collision with a point-supported impurity/edge.

In the present article and after the introduction of our Hamiltonian in Section, 2, we study bound states, scattering coefficients and resonances in the so called regular cases, which are those with $b_d\ne \pm \frac{\hbar^2}{m} \ne b_0$, which is done in Section 3. Next in Section 4, we consider the limit $d\to 0$. In this limit both interactions coincide. The result is again an interaction of the type $a\delta(x)+b\delta'(x)$ and it is quite interesting to evaluate the composition law for the coefficients, which is not linear as one may expect naively. Indeed this composition law establishes the two-dimensional space of couplings $(a,b) \subseteqq \mathbb{R}^2$ as the Borel subgroup of $SL_2({\mathbb R})$, a quite unexpected result.  Section 5 is devoted to study the interesting cases of left-right decoupling limit, i.e., those with $b_d=\pm \frac{\hbar^2}{m}$ and $b_0=\pm \frac{\hbar^2}{m}$. The paper is closed with some concluding remarks.

\section{The Hamiltonian}
Let us consider  a one dimensional free Hamiltonian $H_0$ with a potential of the type  $a\delta(x- d)+ b \delta'(x-d)$. Its Schr\"odinger equation reads,
\begin{equation}
\label{1_1}
-\frac{\hbar^2}{2m}\,\frac{d^2}{dx^2}\,\psi(x)+ 
a\delta(x-d) \psi(x) +b \delta'(x-d) \psi(x)
=E\,\psi(x).
\end{equation}
In order to work with dimensionless quantities, let us introduce new variables and parameters
\begin{equation}
\label{1_2}
x=\frac{\hbar}{mc}\, y,\  
d=\frac{\hbar}{mc}\, q,\   
w_0= \frac{2a}{\hbar c},\   
w_1= \frac{mb}{\hbar^2},\   
\varepsilon= \frac{2E}{m c^2}, \  
\varphi(y)=\psi(x),
\end{equation}
such that \eqref{1_1} becomes
\begin{equation}
\label{1_3}
-\frac{d^2}{d y^2}\,\varphi(y) + 
w_0\delta(y- q) \varphi(y) +2 w_1 \delta'(y- q) \varphi(y)
=\varepsilon \, \varphi(y).
\end{equation}
From now on, we will consider this version of the Schr\"odinger equation instead of \eqref{1_1}.

The point potential we are interested in, ${w_0\delta(y-q)+2 w_1\delta'(y-q)}$, is usually defined via the theory of self-adjoint extensions of symmetric operators of equal deficiency indices \cite{K,K1}, so that the total Hamiltonian $H=H_0+{ w_0\delta(y- q)+2w_1 \delta'(y-q)}$ is self adjoint. The crucial point is finding the domain of wave functions $\varphi(y)$ that makes $H_0$ self adjoint over the domain $\mathbb{R}/\{q\}$ and characterizes the potential ${w_0\delta(y-q)+2 w_1\delta'(y-q)}$. As these functions and their derivatives should have a discontinuity at $y=q$, we have to define the products of the form $\delta({y-q}) \varphi(y)$ and $\delta'({y-q}) \varphi(y)$ in \eqref{1_3}.  These can be done in several ways \cite{Za1,Za2,Za3}, but we choose the following:
\begin{eqnarray}
 \label{1_222}
\delta({y-q})  \varphi(y)\!\!&\!\!=\!\!&\!\! \frac{\varphi({q}^+)+\varphi({q}^-)}{2}\,\delta({y-q})\,,  \\ [1ex] 
\delta'({y-q}) \varphi(y) \!\!&\!\!=\!\!&\!\!  \frac{\varphi({q}^+)+\varphi({q}^-)}{2}\,\delta'({y-q})-\frac{\varphi'({q}^+)+\varphi'({q}^-)}{2}\,\delta({y-q})\,,
 \label{1_222_2}
\end{eqnarray}
where $f({q}^+)$ and $f({q}^-)$ are the right and left limits of the function $f({y})$ as ${y}\to {q}$, respectively. The Schr\"odinger equation (\ref{1_3}) should be viewed as a relation between distributions.

In order to obtain a self-adjoint determination of the Hamiltonian $H=H_0+w_0\delta(y-q)+2w_1\delta'(y-q)$, we have to find a self adjoint extension  of $H_0$. In order to do it, we have to find a domain on which this extension acts. This domain is given by a space of square integrable functions satisfying certain assumptions including matching conditions at the point $q$ that affects to the value of wave functions and their derivatives at $q$ \cite{AL,K}. In particular, this implies that both wave functions and derivatives cannot be continuous at $q$, so that equations  \eqref{1_222} and \eqref{1_222_2} make sense.

The functions in the domain of the Hamiltonian $H$ are functions in the Sobolev space\footnote{This is the space of absolutely continuous functions $f(y)$ with absolutely continuous derivative $f'(y)$, both having arbitrary discontinuities at $q$, such that the Lebesgue integral given by
$$
\int_{-\infty}^\infty \{|f(y)|^2+|f''(y)|^2\}\,dy
$$
converges.} $W^2_2({\mathbb R}\setminus\{q\})$ such that at $q$ satisfy the following matching conditions\footnote{This is true for $w_1\ne\pm 1$, while for $w_1=\pm 1$, we have to define the matching conditions in another way \cite{K}. We shall concentrate now in the regular cases, but the decoupling limit will be also considered later on in Section~\ref{exceptional}.}:
\begin{equation}\label{1_4}
\left( \begin{array}{c} 
\varphi(q^+) \\[1ex] 
\varphi'(q^+)  \end{array}  \right) =  
\left( \begin{array}{cc}  
\displaystyle\frac{1+w_1}{1-w_1}  &  0 
\\[1ex]  
\displaystyle
\frac{  w_0}{1-w_1^2}  &  \displaystyle \frac{1-w_1}{1+w_1} \end{array}  \right) \left( \begin{array}{c} 
\varphi(q^-)  \\[1ex] \varphi'(q^-)  \end{array}  \right)\,.
\end{equation} 
These results could be, in principle, extended to interactions of the type $\sum_{i=1}^N  a_i\delta(y-q_i)+2 b_i\delta'(y-q_i)$, where $N$ could be either finite or infinite \cite{K}. In the present paper, we  assume that $N=2$ as above, so that the total Hamiltonian takes the form:
\begin{equation}\label{1_5}
H=H_0+V+W\,, \qquad H_0= -\frac{d^2}{d y^2}\,,
\end{equation}
with
\begin{equation}\label{1_6}
V= v_0\delta(y)+2v_1\delta'(y)\, ,\qquad  W= w_0\delta(y-q)+2w_1\delta(y-q)\, ,
\end{equation}
where $v_0=\frac{2A}{\hbar c}$, $v_1=\frac{m B}{\hbar^2}$, and $A$, $B$ are respectively the $\delta$ and $\delta^\prime$ couplings of the pair placed at the origin in the original variables. While $V$ is supported at the origin $y=0$, $W$ is supported at a point $q$ that we assume positive, $q>0$. The potential $V+W$ is physically relevant as is related to the Casimir effect \cite{MM}. 
The corresponding dimensionless Schr\"odinger equation is
\begin{equation}\label{1_9}
-\frac{d^2 \varphi(y)}{d y^2} + \left[
v_0\delta(y)+2v_1\delta'(y)+
w_0\delta(y- q) +2w_1 \delta'(y- q) \right] \varphi(y)
=\varepsilon \, \varphi(y).
\end{equation}
One of the motivations of the present work was the study of the effect resulting of taking the limit in (\ref{1_9}) as $q\to 0$, i.e., the effect of the superposition of the two point potentials $V$ and $W$ at the same point. We shall see that, as a result of the limit, we obtain a point potential of the type $u_0\delta (y)+{2}u_1\delta'(y)$, where $u_0$ and $u_1$ are not the sums  $v_0+w_0$ and $v_1+w_1$, but instead another kind of law, which has a group structure as the Borel subgroup of $SL_2({\mathbb R})$. Some further results will be discussed.

\section{Matching conditions and scattering coefficients}

\begin{figure}[ht]
\centering
\includegraphics[width=0.42\textwidth]{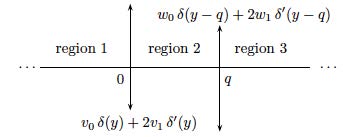}
\caption{\small Schematic representation of the physical problem under study, given by \eqref{1_9}, before considering the limit $q\to 0$: a potential with a couple of singular point interactions described by a linear combination of Dirac delta distributions and their derivatives.
\label{fig1}}
\end{figure}

In order to analyze the solution of \eqref{1_9}, let us split the real line into three regions, as is shown in Figure \ref{fig1}: region 1 is the half line with $y<0$, region 2 is the segment $0<y<q$ and region 3 is the half line $y>q$. Supported on $y=0$ and on $y=q$, we have the point potentials $V$ and $W$, as in (\ref{1_6}), respectively. In these regions the values of the solution and their derivatives are those of the free Schr\"odinger equation,  given by:
\begin{equation}\label{1_7}
\varphi_j(y)=A_j\,e^{-ik y}+B_j\,e^{ik y}\,,\quad \varphi'_j(y)=-ik(A_j\,e^{-iky}-B_j\,e^{iky})\,,\quad j=1,2,3\,,\quad k^2=\varepsilon>0.
\end{equation}
Since the point potential $V$ is defined by matching conditions like those in (\ref{1_4}), we have that
\begin{equation}\label{1_8}
\left(\begin{array}{c} A_2+B_2\\  -ik(A_2-B_2)\end{array} \right) = M_v   \left(\begin{array}{c} A_1+B_1\\  -ik(A_1-B_1)  \end{array}   \right),\quad 
M_v=\left( \begin{array}{cc} \displaystyle \frac{1+v_1}{1-v_1}  &  0 \\[1ex] 
\displaystyle \frac{v_0}{1-v_1^2}  & \displaystyle \frac{1-v_1}{1+v_1} \end{array}  \right)\,.
\end{equation}
Defining the following matrix:
\begin{equation}\label{1_999}
K=\left(\begin{array}{cc} 1  & 1 \\  -ik  &  ik   \end{array}   \right)\,,
\end{equation}
equation (\ref{1_8}) becomes:
\begin{equation}\label{1_10}
\left( \begin{array}{c} A_2  \\   B_2  \end{array}  \right)= K^{-1}\, M_v\, K \left( \begin{array}{c} A_1  \\  B_1  \end{array}  \right)\,.
\end{equation}
This expression (\ref{1_10}) gives the matching conditions at the point $y=0$. At the point $y=q$ the same procedure works after appropriate translation
\begin{equation}\label{1_10bis}
\left( \begin{array}{c} A_3  \\   B_3  \end{array}  \right)= Q^{-1}\,K^{-1}\,M_w\,K\,Q \left( \begin{array}{c} A_2  \\   B_2  \end{array}  \right)\,.
\end{equation}
and we finally obtain:
\begin{eqnarray}\label{1_13}
\left( \begin{array}{c} A_3  \\   B_3  \end{array}   \right)
= Q^{-1}\,K^{-1}\,M_w\,K\,Q\,K^{-1}\, M_v\, K \left( \begin{array}{c} A_1  \\   B_1  \end{array}  \right)  
=T_q\, \left( \begin{array}{c} A_1  \\   B_1  \end{array}  \right)\,,
\end{eqnarray}
where the definition of $T_q$ is obvious, while $Q$ and $M_w$ are respectively:
\begin{eqnarray}\label{1_12}
Q= \left( \begin{array}{cc} e^{-iqk} & 0 \\   0  &   e^{iqk}  \end{array}   \right)\,,\qquad 
M_w=\left( \begin{array}{cc}  
\displaystyle\frac{1+w_1}{1-w_1}  &  0 \\[1ex] 
\displaystyle\frac{w_0}{1-w_1^2}  & \displaystyle \frac{1-w_1}{1+w_1} \end{array}  \right)\,.
\end{eqnarray}
The $T_q$-matrix in (\ref{1_13}) relates the asymptotic behaviour of the two linearly independent Jost scattering solutions at $x<< 0$ with their counterparts at $x>> 0$, see e.g. \cite{Boya}. The $T_q$-matrix elements satisfy the identities:
\[
{\rm det}T_q= T_q^{11}T_q^{22}-T_q^{12}T_q^{12}=1 \, \, \, , \, \, \, T_q^{11}=\bar{T}^{22}_q \, \, \, , \, \, \, T_q^{12}=\bar{T}^{21}_q 
\]
such that $T_q$ is in general an element of the group $SL_2({\mathbb C})$. From the $T_q$-matrix one obtains the scattering matrix $S_q$ by an standard
procedure. Reshuffling the linear system (\ref{1_13}) in the form
\begin{equation}\label{1_24}
\left( \begin{array}{c} A_1  \\   
B_3  \end{array}   \right)=
S_q \left( \begin{array}{c} A_3 \\   
B_1 \end{array} \right)
\end{equation}
one easily checks that
\begin{equation}\label{1_251}
S_q=\frac{1}{T_q^{11}}\left(\begin{array}{cc} 1 & -T_q^{12} \\ T_q^{21} & {\rm det}T_q \end{array}\right) \, \, \, , \, \, \, S_q^\dagger =\frac{1}{\bar{T}_q^{11}}\left(\begin{array}{cc} 1 & \bar{T}_q^{21} \\ -\bar{T}_q^{12} & {\rm det}T_q \end{array}\right) \, \, , \, \, S_q^\dagger S_q=\left(\begin{array}{cc} 1 & 0 \\ 0 & 1\end{array}\right) \, \, ,
\end{equation}
i.e., the scattering $S_q$-matrix arises as a $2\times 2$ unitary matrix: $S_q^\dagger S_q=S_qS_q^\dagger=1$. The transmission and reflection coefficients provide the usual form of writing the $S_q$-matrix elements:
\begin{equation}\label{1_252}
S_q=\left(\begin{array}{cc} t(k) & r_R(k) \\ r_L(k) & t(k) \end{array}\right) \, \, .
\end{equation}
$t(k)$ is the amplitude of the transmitted waves coming from the far left or from the far right. Time-reversal invariant potentials 
give rise to identical transmission coefficients: $t_R(k;q)=t_L(k;q)=t(k;q)$. However the reflection amplitudes are different for incoming waves either from the far right or the far left such that $r_R(k;q) \neq r_L(k;q)$ because the potential is not parity invariant. Comparison of equations (\ref{1_251}) and (\ref{1_252}) shows that:
\begin{equation}\label{1_253}
t(k;q)=\frac{1}{T_q^{11}(k)} \, \, \, , \, \, \, r_R(k;q)=-\frac{T_q^{12}(k)}{T_q^{11}(k)} \, \, , \, \, r_L(k,q)=\frac{T_q^{21}(k)}{T_q^{11}(k)} \, .
\end{equation}
Transmission and reflection coefficients were calculated in \cite{MM} when the two $\delta$-$\delta^\prime$ point interactions were
symetrically located with respect to the origin. For the arrangement chosen in this paper we find from formulas (\ref{1_253}) :
\begin{eqnarray}
r_L(k;q)  \!\!&\!\!=\!\!&\!\!   - \frac{e^{-2 i q k} \left(2 k \left(v_1^2+1\right)+i v_0\right) \left(4 k w_1-i
   w_0\right)+\left(4 k v_1-i v_0\right) \left(2 k \left(w_1^2+1\right)-i
   w_0\right)}{\Delta(k)} ,
\label{1_30} \\[1ex]
r_R(k;q) \!\!&\!\!=\!\!&\!\!  \frac{e^{2 i q k} \left(2 k \left(v_1^2+1\right)-i v_0\right) \left(4 k w_1+i
   w_0\right)+\left(4 k v_1+i v_0\right) \left(2 k \left(w_1^2+1\right)+i
   w_0\right)}{\Delta(k)}, \label{1_31}\\
   [1ex]
t(k;q) \!\!&\!\!=\!\!&\!\!  \frac{4k^2(1-v_1^2)(1-w_1^2)}{\Delta(k)}  \\
[1ex] \Delta(k) \! \! & \!\!= \!\! & \! \! e^{2ikq}(v_0+4ikv_1)(w_0-4ikw_1)+(2kv_1^2+2k+iv_0) (2kw_1^2+2k+iw_0), \label{1_32} 
\end{eqnarray}
where $\Delta(k)$ is a function of $k$ and the other parameters of the problem.

\subsection{Bound/antibound states and resonances}\label{sectionboundetc}

Complex zeroes of 
\begin{equation}\label{1_25}
 T_q^{11}(k)=\frac{\Delta(k)}{4 k^2(1-v_1^2)(1-w_1^2)}\, \, ,
\end{equation}
which are poles of the $S_q$-matrix in the $k$-complex plane, give rise to bound or antibound states if are located on the purely imaginary,
respectively positive or negative half-axis. Complex zeroes coming in pairs having opposite real part and identical imaginary part correspond to resonances.

\begin{figure}[ht]
\centering
\includegraphics[width=0.4\textwidth]{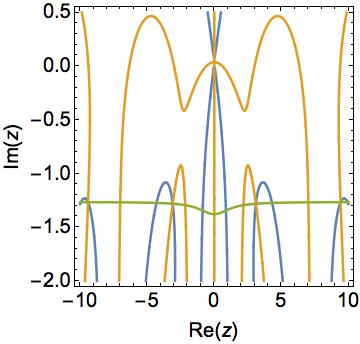}
\quad
\includegraphics[width=0.5679\textwidth]{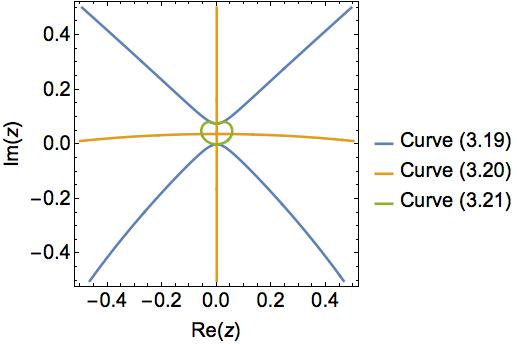}
\caption{\small 
On the left, for some values of the parameters ($v_0=-2$, $v_1=4$, $w_0=-1$, $w_1=3$ and $q=1/2$), the typical behavior of the curves  \eqref{2real_lambert1}-\eqref{compatibility} is plotted. The intersections between the blue and orange curves (the values of $z$ such that the real and imaginary parts of $\Delta(z)$ are simultaneously zero) are seen as discrete points, placed on the compatibility condition curve (green).
Remark that there is an infinity number of such solutions on the lower part of the complex plane, appearing in pairs $z=\pm z_{r,n}-i z_{i,n},\, z_{r,n}, z_{i,n}> 0, n=1,2,\dots$, corresponding accordingly to resonances in the spectrum; in addition, there is also one imaginary solution with positive imaginary part $z_0= i z_{i,0}$.
On the right, a zoom of the same plot near the origin is shown. One may observe a bound state in the upper intersection. The lower intersection, however, happens exactly at the origin and it is not a meaningful physical solution.}
\label{lambert_curves}
\end{figure}

The relevant physical information is encoded in the analysis of complex zeroes of $T_q^{11}(k)$, or equivalently
of $\Delta(k)$. In order to simplify this analysis, it is useful to make the following changes in the parameters in
\eqref{1_25}:
\begin{equation}\label{changeparameters}
2kq=z, \quad
\sigma=\frac{q v_0}{1+v_1^2},\quad
\tau=\frac{q w_0}{1+w_1^2},\quad
v=\frac{v_1}{1+v_1^2},\quad
w= \frac{w_1}{1+w_1^2}.
\end{equation}
Then, $\Delta(k)=0$ is equivalent to 
\begin{equation}\label{Delta=zero}
e^{iz}=-\left(\frac{z+i\sigma}{\sigma+2iv z} \right)
\left(\frac{z+i\tau}{\tau-2iwz}\right) ,\quad z=z_r+iz_i,\ z_r,z_i\in{\mathbb R}.
\end{equation}
Equation \eqref{Delta=zero} can be considered as a complex or two-dimensional generalization of the so-called {\em generalized Lambert equation} (see \cite{Scott1,Scott2} and references quoted therein). From this complex Lambert equation we can obtain two real equations, corresponding to the real and imaginary parts of \eqref{Delta=zero}:
\begin{eqnarray}
\label{2real_lambert1}
\hskip-0.9cm
 \left[ 4 v w z_r^2 -(2vz_i-\sigma)(2wz_i+\tau)\right] \cos z_r -2 z_r
   (\tau  v+4 v w z_i-\sigma  w) \sin z_r 
  \!\! &\!\!=\!\!&\!\!  \left[ (z_i+\sigma )( z_i+\tau) -z_r^2 \right] e^{z_i} .
\\
\hskip-0.9cm
2 z_r (\tau  v+4 v w z_i-\sigma  w) \cos z_r +
   \left[  4 v w z_r^2 -(2vz_i-\sigma)(2wz_i+\tau) \right]  \sin z_r 
    \!\! &\!\!=\!\!&\!\! -z_r(2z_i+ \sigma +\tau) e^{z_i}.
\label{2real_lambert2}
\end{eqnarray}
These expressions for real and imaginary parts of $\Delta(k)=0$ are rather complicated. A simpler compatibility condition is given by
\begin{equation}
\label{compatibility}
e^{2 z_i}=\frac{(4 v^2 z_r^2+\left(2 v z_i- \sigma )^2\right)   \left(4 w^2 z_r^2+ (2 w z_i+\tau )^2\right)}{
\left( z_r^2+ (z_i+ \sigma )^2\right)
   \left(z_r^2+ (z_i+ \tau )^2\right)} ,
\end{equation}
which is, indeed, a generalization of the Lambert equation that includes two real variables. The solutions we are looking for must satisfy \eqref{compatibility}, although \eqref{2real_lambert1}--\eqref{2real_lambert2} are more restrictive.
First of all, we can easily check that a simple solution of the system \eqref{2real_lambert1}--\eqref{2real_lambert2} is $z_r=0,\, z_i=0$, but this corresponds to $k=0$, which is a pole of 
$T^{11}_q(k)$.
In addition, it is obvious that the system is symmetric on the variable $z_r$, although has no symmetry on $z_i$.
The typical behavior of the solutions of \eqref{2real_lambert1}--\eqref{compatibility} is shown in Figure~\ref{lambert_curves}.

\section{The addition law at the $q=0$ limit}
\paragraph{Proposition 1.}\textit{ The $T_q$ matrix for two-pairs of $\delta$-$\delta^\prime$ interactions displaced from each other a distance q is determined in formula \eqref{1_13}. In the limit $q\rightarrow 0$ $T_q$ becomes:
\begin{equation}
\left( \begin{array}{c} A_3  \\   B_3  \end{array}  \right)= K^{-1}\, M_w\cdot M_v\, K \left( \begin{array}{c} A_1  \\  B_1  \end{array}  \right)= K^{-1}\, M_u\, K \left( \begin{array}{c} A_1  \\  B_1  \end{array}  \right)\,.
\end{equation}
That is to say, the superposition of these two $\delta$-$\delta^\prime$ is equivalent to a single $\delta$-$\delta^\prime$ where the matching conditions are characterized by the Kurasov matrix 
\begin{equation}
M_u=\left(\begin{array}{cc}\frac{1+u_1}{1-u_1} & 0 \\ \frac{u_0}{1-u_1^2} & \frac{1-u_1}{1+u_1} \end{array}\right)\quad {\rm where}\quad u_1= \frac{v_1+w_1}{1+v_1w_1}\,,\quad 
u_0=\frac{v_0(1-w_1)^2+w_0(1+v_1)^2}{(1+v_1w_1)^2}\,.
\end{equation}}

\textit{Proof.} In the $q=0$ limit, the transfer matrix is $\lim_{q\to 0}T_q=T_0$, while $Q$ obviously becomes the identity matrix. Since $M_w$ and $M_v$ do not depend on $q$, the $q=0$ limit of \eqref{1_13} reads:
\begin{equation}\label{1_14}
\left( \begin{array}{c} A_3  \\   B_3  \end{array}   \right)= K^{-1}\,M_w\,M_v\,K  \left( \begin{array}{c} A_1  \\   B_1  \end{array}  \right)=
T_0 \left( \begin{array}{c} A_1  \\   B_1  \end{array}  \right)  .
\end{equation}
One may expect that the product of two Kurasov matrices $M_w\,M_v$ is another Kurasov matrix. This assumption means that
\begin{equation}\label{1_15}
M_w\,M_v=
\left( \begin{array}{cc}  \displaystyle
\frac{1+w_1}{1-w_1}  &  0 \\[1ex] 
\displaystyle \frac{w_0}{1-w_1^2}  & \displaystyle \frac{1-w_1}{1+w_1} \end{array}  \right) 
\left( \begin{array}{cc}  \displaystyle \frac{1+v_1}{1-v_1}  &  0 \\[1ex] 
\displaystyle \frac{v_0}{1-v_1^2}  & \displaystyle\frac{1-v_1}{1+v_1} \end{array}  \right)  
=
\left( \begin{array}{cc} \displaystyle \frac{1+u_1}{1-u_1}  &  0 \\[1ex] 
\displaystyle\frac{u_0}{1-u_1^2}  & \displaystyle\frac{1-u_1}{1+u_1} \end{array}  \right)=M_u .
\end{equation}
A lengthy but straightforward calculation shows that this is the case with the \lq\lq composite\rq\rq couplings:
\begin{equation}\label{1_16}
u_1= \frac{v_1+w_1}{1+v_1w_1}\,,\qquad 
u_0=\frac{v_0(1-w_1)^2+w_0(1+v_1)^2}{(1+v_1w_1)^2}\, ,
\end{equation}
Thus, the collapsed two $\delta$-$\delta^\prime$ pairs are tantamount to a single point potential of the form $u_0\delta(y)+2u_1\delta'(y)$
defined by the transfer matrix $T_0=K^{-1}M_u K$, q. e. d.

One would have expected another result such as an additive rule of the type $u_0=v_0+w_0$ and $u_1=v_1+w_1$, due to the form of point potentials that converge. In fact for zero $\delta^\prime$ couplings we would obtain:
\[
\left( \begin{array}{cc}  \displaystyle
1  &  0 \\[1ex] 
\displaystyle w_0  & \displaystyle 1 \end{array}  \right) 
\left( \begin{array}{cc}  \displaystyle 1  &  0 \\[1ex] 
\displaystyle v_0  & 1 \end{array}  \right)  
=
\left( \begin{array}{cc} \displaystyle 1 &  0 \\[1ex] 
\displaystyle w_0+v_0  & \displaystyle 1 \end{array}  \right)
\]
and the fusion of two $\delta$ point potentials is a pure abelian process.
Nonetheless, the addition law (\ref{1_16}) is quite interesting. First of all, the first equation in (\ref{1_16}) resembles the addition law of velocities in special relativity. However, the second one looks more complicated.  In any case, one natural question we may pose is if the composition law (\ref{1_16}) has a structure of a group and, if this is the case, which group this can be. 

Checking the group structure is quite simple. The product of two matrices $M_v$ and $M_w$ as in \eqref{1_15}
gives another matrix with identical structure. This gives the product law.
The associativity comes from the associativity of the product of matrices. 
The identity of the group is the $2\times 2$ identity matrix $\mathbb I$, which corresponds to take both parameters equal to zero in \eqref{1_4}, $M_0={\mathbb I}$. The inverse of $M_v$ is a matrix $M_s$ so that 
$M_v\,M_s = {\mathbb I}$.
The calculation is straightforward. For $M_v$ as defined in (\ref{1_8}) one finds
\begin{equation}
M_s=\left( \begin{array}{cc}  \displaystyle\frac{1-v_1}{1+v_1}  & 0 \\[1ex]  
\displaystyle\frac{-v_0}{1-v_1^2}  &  \displaystyle\frac{1+v_1}{1-v_1}  \end{array}  \right),
\end{equation} 
which shows that taking the inverse in our group of matrices is equivalent to the transformation $(v_0,v_1) \to (-v_0,-v_1)$. We may denote the inverse of $M_v$ as $M_{-v}$. Thus, the structure of group in our set of matrices has been confirmed. Furthermore, there are two other remarkable properties of matrices in this group:

\smallskip
(i) Non commutativity:

\begin{equation}
M_v M_w-M_w M_v= 
\left( \begin{array}{cc} 0 & 0 \\[1ex]  
\displaystyle\frac{4(v_0w_1-w_0v_1)}{(1-v_1^2)(1-w_1^2)} & 0   \end{array}  \right)\,.
\end{equation}

\smallskip

\begin{figure}[ht]
\centering
\includegraphics[width=0.35\textwidth]{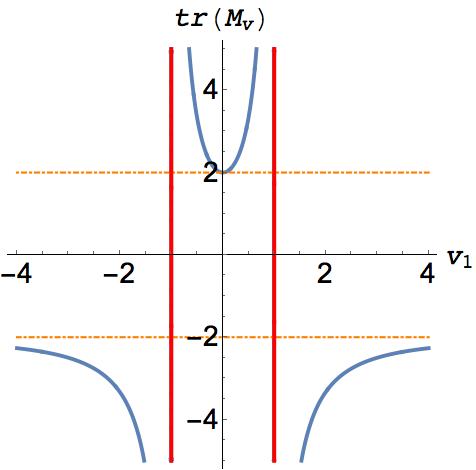}
\caption{\small Plot of the trace of a generic matching matrix $M_v$, given by equation \eqref{traza}.
\label{fig2}}
\end{figure}

(ii) The trace of $M_v$ is independent of $v_0$:
\begin{equation}
\label{traza}
{\rm tr}\, M_v=2\,\frac{1+v_1^2}{1-v_1^2}.
\end{equation}
This trace has two singular points $v_1=\pm 1$,  emerging from Kurasov's analysis, and corresponding to the decoupling limits which are not included in the type of matrices that we are considering in the present section (for an analysis of the decoupling or completely opaque limits see Section~\ref{exceptional}). It is positive on the interval $-1<v_1<1$, reaching the minimum value 2 for $v_1=0$. On the half lines $-\infty<v_1<-1$ and $1< v_1<\infty$ the trace is negative and its modulus is bigger than 2. Its limits when $v_1\to \pm \infty$ are equal to $-2$. 
A plot of the function \eqref{traza} can be seen on Figure~\ref{fig2}.

\subsection{The Borel subgroup of $SL_2({\mathbb R})$ and its Lie algebra}
\paragraph{Proposition 2.}\textit{ The Kurasov Matrices form a group where the matrix product $M_v\cdot M_w=M_u$ is the group law:
\begin{equation}
u_1= \frac{v_1+w_1}{1+v_1w_1}\,,\qquad 
u_0=\frac{v_0(1-w_1)^2+w_0(1+v_1)^2}{(1+v_1w_1)^2}\,.
\end{equation}
The group of Kurasov Matrices is the Borel subgroup of $SL_2(\mathbb{R})$}

\textit{Proof.} All the previous information shows that the set of $M_v$ matrices is the subgroup of $SL_2(\mathbb{R})$ of all the lower triangular matrices such that $\vert{\rm tr}(g)\vert\geq2$. This is the Borel subgroup of $SL_2(\mathbb{R})$. 

The Lie algebra of the Borel subgroup of $SL_2(\mathbb{R})$ is the minimal parabolic subalgebra of $\mathfrak{sl}_2(\mathbb{R})$. The Cartan basis for $\mathfrak{sl}_2(\mathbb{R})$ is given by the following $2\times 2$ matrices:
\begin{equation}
e=\left(
\begin{array}{cc}
 0 & 1 \\
 0 & 0 \\
\end{array}
\right),\qquad 
f=\left(
\begin{array}{cc}
 0 & 0 \\
 1 & 0 \\
\end{array}
\right),\qquad 
h=\left(
\begin{array}{cc}
 1 & 0 \\
 0 & -1 \\
\end{array}
\right).
\end{equation}
Their commutation relations are 
\begin{equation}
[e,f]=h\,, \quad [h,f]=-2 f\,,\quad [h,e]=2 e. 
\end{equation}
The Borel subalgebra of a given Lie algebra $\mathfrak{g}$ with Cartan subalgebra $\mathfrak{h}$ is the direct sum of the Cartan subalgebra  with a given ordering and the weight spaces of $\mathfrak{g}$ with negative weight (it is equivalent to take positive weights; this would just change upper diagonal matrices by lower diagonal matrices. In any case, both algebras, or equivalently both groups, are isomorphic). Therefore in the case of $\mathfrak{sl}_2(\mathbb{R})$ the Borel subalgebra is
\begin{equation}
\mathfrak{g}_{K(x_0)}=\langle h, f\rangle \,.
\end{equation}
The elements of the two disconnected components of the group are obtained by exponentiation:
\begin{equation}
e^{\alpha h+ \beta f}=\left(
\begin{array}{cc}
 e^{\alpha } & 0 \\     
\displaystyle\beta\ \frac{  \sinh \alpha}{\alpha } & e^{-\alpha } \\
\end{array}
\right) \,,  \qquad 
-e^{\alpha h+ \beta f}=\left(
\begin{array}{cc}
 -e^{\alpha } & 0 \\[1ex]
\displaystyle - \beta\ \frac{  \sinh \alpha}{\alpha } & -e^{-\alpha } \\
\end{array}
\right)\,.
\end{equation}
For the component of the elements connected with the identity, the couplings are easily obtained in terms of the group parameters:
\begin{equation}
e^{\alpha h+ \beta f}=M_{v} \quad \Longrightarrow \quad
v_0= 2\, \frac{\beta}{\alpha}\, {\tanh}\frac{\alpha}{2}  , \quad v_1={\tanh}\frac{\alpha}{2} \,.
\end{equation}
On the other hand, for the elements in the other connected component, we have:
\begin{equation}
-e^{\alpha h+ \beta f}=M_{v} \quad \Longrightarrow \quad
v_0= 2\, \frac{\beta}{\alpha}\, {\coth}\frac{\alpha}{2} ,  
\quad v_1={\coth}\frac{\alpha}{2}\,.
\end{equation}
The composition law of the group can be expressed in terms of these exponentials. In fact,
\begin{equation}
e^{\alpha_1 h+ \beta_1 f} \ e^{\alpha_2 h+ \beta_2 f}= e^{\alpha h+ \beta f}
\end{equation}
with
\begin{equation}
\alpha=\alpha_1+\alpha_2\,,\qquad 
\beta= \frac{ \left(\alpha_1+\alpha _2\right)
   \left(e^{\alpha _2}
   \alpha _2 \beta _1 \sinh
   \left(\alpha _1\right)+e^{-\alpha _1}\alpha _1
   \beta _2 \sinh \left(\alpha
   _2\right)\right)}{\alpha_1\alpha _2 \sinh(\alpha_1+\alpha_2)}.
\end{equation}
This concludes the discussion on the group structure. Q. E. D

\subsection{Reflection and transmission coefficients due to a single $\delta$-$\delta^\prime$ interaction at the origin}
\paragraph{Proposition 3.}\textit{ The $q=0$ limit of the scattering coefficients of two a priori separated pairs of $\delta$-$\delta^\prime$ interactions exactly coincide with the scattering coefficients produced by a single $\delta$-$\delta^\prime$ pair with couplings determined from the $M_u=M_v\cdot M_w$ Kurasov matrix. In terms of the scattering matrix we can write:
\begin{equation}
\lim_{q\rightarrow 0} S_q(M_v,M_w)=S(M_v\cdot M_w)=S(M_u).
\end{equation}
Hence the scattering matrices produced by two superimposed pairs of $\delta$-$\delta^\prime$ interactions give rise to a representation
of the non-abelian composition law of Kurasov matrices.
}

\textit{Proof.} The linear map between asymptotic scattering Jost solutions respectively in the far right and the far left,
\[
\left(\begin{array}{c} A_2 \\ B_2\end{array}\right)=Z\left(\begin{array}{c} A_1 \\ B_1 \end{array}\right),
\]
produced by a single point interaction $V(x)=u_0 \delta(x)+2u_1\delta^\prime(x)$ is provided by a similarity transformation of the Kurasov matrix :
\[
Z=K^{-1} M_u K \quad , \quad  M_u=\left(\begin{array}{cc} 
\displaystyle\frac{1+u_1}{1-u_1} & 0 \\ 
\displaystyle\frac{u_0}{1-u_1^2} & \displaystyle\frac{1-u_1}{1+u_1}\end{array}\right) \, \, .
\]
We obtain for the $Z$-matrix elements
\begin{equation}\label{1_254}
Z^{11}(k)=\frac{2k(1+u_1^2)+i u_0}{2 k(1-u_1^2)}=\bar{Z}^{22}(k) \quad , \quad Z^{12}(k)=\frac{4ku_1+iu_0}{2k(1-u_1^2)}=\bar{Z}^{21},
\end{equation}
where the bar stands for complex conjugation. From this information one reads the scattering coefficients
\begin{eqnarray}
r_R(k)&=&-\frac{Z^{12}(k)}{Z^{11}(k)}=-\frac{4ku_1+iu_0}{2k(u_1+1)+iu_0}\,, \qquad r_L(k)=\frac{Z^{21}(k)}{Z^{11}(k)}= \frac{4ku_1-iu_0}{2k(u_1^2+1)+iu_0},\\
t(k)&=&\frac{1}{Z^{11}(k)}=-\frac{2k(u_1^2-1)}{2k(u_1^2+1)+iu_0},\, \label{1_38}
\end{eqnarray}
in perfect agreement with the results in  \cite{GNN}.

Finally, we compare these scattering coefficients with those obtained  from the $T_0$-matrix that describe the map between asymptotic
Jost solutions when the distance $q$ from one $\delta$-$\delta^\prime$ point interaction to the other tends to zero. The outcome is:
\begin{eqnarray}
&& \hspace{-1.2cm}r_L(k;0) =-\frac{T_0^{12}(k)}{T_0^{11}(k)}=  
-\frac{v_1 \left(\left(v_1+2\right) w_0+4 i k\right)+4 i k v_1 w_1
   \left(v_1+w_1\right)+4 i k w_1+v_0 \left(w_1-1\right){}^2+w_0}{-2 i k w_1
   \left(\left(v_1^2+1\right) w_1+4 v_1\right)-2 i k
   \left(v_1^2+1\right)+\left(v_1+1\right){}^2 w_0+v_0 \left(w_1-1\right){}^2}, \label{1_35} 
\\[1ex]
&& \hspace{-1.2cm}r_R(k;0)=\frac{T_0^{21}(k)}{T_0^{11}(k)} =  
 \frac{v_1 \left(-\left(v_1+2\right) w_0+4 i k\right)+4 i k v_1 w_1
   \left(v_1+w_1\right)+4 i k w_1-v_0 \left(w_1-1\right){}^2-w_0}{-2 i k w_1
   \left(\left(v_1^2+1\right) w_1+4 v_1\right)-2 i k
   \left(v_1^2+1\right)+\left(v_1+1\right){}^2 w_0+v_0 \left(w_1-1\right){}^2},  \label{1_34} 
\\[1ex]
&& \hspace{-1.2cm}t(k;0)=\frac{1}{T_0^{11}(k)} =  
\frac{2 k \left(v_1^2-1\right) \left(w_1^2-1\right)}{2 k w_1
   \left(\left(v_1^2+1\right) w_1+4 v_1\right)+2 k \left(v_1^2+1\right)+i
   \left(v_1+1\right){}^2 w_0+i v_0 \left(w_1-1\right){}^2}, \label{1_362}
\end{eqnarray}
to check that the addition law
\[
u_1= \frac{v_1+w_1}{1+v_1w_1}\,,\qquad 
u_0=\frac{v_0(1-w_1)^2+w_0(1+v_1)^2}{(1+v_1w_1)^2}\, ,
\]
also works for the scattering coefficients, \textit{q.e.d}.

\section{The left-right decoupling values of the $\delta^\prime$ couplings}\label{exceptional}

We have seen that there are singularities of the Kurasov matrix at the critical points, $v_1=\pm 1$ and $w_1=\pm 1$. One may expect that these critical cases where the transmission coefficients are zero do not contribute to the group structure that we have previously analyzed.  The completely opaque potentials as given in equations (\ref{1_6}) have clearly one of the following four forms:
\begin{equation}\label{1_46}
V=v_0\delta(x)\pm 2\delta'(x)  
\quad {\rm and/or}\quad
 W=w_0\delta(x-q) \pm 2\delta'(x-q)\,.
\end{equation}
In order to define these potentials, we obviously cannot use matching conditions of the form (\ref{1_4}) that are singular. Instead, we shall impose \cite{K}: 
\begin{enumerate}
\item 
For $v_1=1$:
\begin{equation}\label{1_47}
\varphi(0^-)=0\,, \quad \varphi'(0^+)= \frac{v_0}{4}\, \varphi(0^+).
\end{equation} 
\item
For $w_1=1$:
\begin{equation}\label{1_47bis}
 \varphi(q^-)=0\,,\quad \varphi'(q^+)= \frac{w_0}{4}\,\varphi(q^+)\,.
\end{equation} 

Thus, we set Dirichlet boundary conditions on the left of the two points $x=0$ and $x=q$, whereas Robin boundary conditions are chosen on the right hand side of $x=0$ and $x=q$, see \cite{MM}. In the two remaining cases the matching conditions are the mirror images of the previous ones.

\item 
For $v_1=-1$:
\begin{equation}\label{1_48}
\varphi(0^+)=0\,, \quad \varphi'(0^-)= -\frac{v_0}{4}\, \varphi(0^-).
\end{equation}

\item 
For $w_1=-1$:
\begin{equation}\label{1_48bis}
 \varphi(q^+)=0\,,\quad \varphi'(q^-)= -\frac{w_0}{4}\,\varphi(q^-)\,.
\end{equation} 

\end{enumerate}
We remain, however, in the physical situation depicted in Figure~\ref{fig1}, were the plane waves and their derivatives are written as in (\ref{1_7}). There are eight possible configurations involving at least one decoupling configuration of the couplings: either the two $\delta$-$\delta^\prime$ interactions build opaque walls both at $x=0$ and $x=q$, or, there is no transmission only at one point. We discuss first the cases when the two $\delta^\prime$ couplings take the decoupling limit.

\subsection{Two $\delta^\prime$ couplings in the decoupling limit}

There are four cases in which we have decoupling situations both at $x=0$ and $x=q$. Let us consider them separately.
\smallskip

\noindent
\underline{Case 1}: $v_1=1,w_1=1$. 

\noindent
The boundary conditions are given by \eqref{1_47} and \eqref{1_47bis}. 
Let us consider the situation on the interval $[0,q]$. 
For $x=0$, $\varphi'(0^+)= \frac{v_0}{4}\, \varphi(0^+)$ is written as:
\begin{equation}\label{1_49}
-ik(A_2-B_2)=\frac{v_0}{4}\, (A_2+B_2) \Longrightarrow \frac{A_2}{B_2}=
-\frac{v_0-4ik}{v_0+4ik}=-\exp\left( -2i\arctan \frac{4k}{v_0} \right) .
\end{equation}
For $x=q$, we write $\varphi(q^-)=0$ as
\begin{equation}\label{1_50}
A_2\,e^{-ikq}+B_2\,e^{ikq}=0 \Longrightarrow  \frac{A_2}{B_2}= -e^{2ikq}\,.
\end{equation}
Then, we have a transcendental equation, which can be written either in the form of a generalized Lambert equation:
\begin{equation}\label{1_51}
e^{2ikq}= \frac{v_0-4ik}{v_0+4ik}\,,
\end{equation}
or in terms of the function $\arctan$ as
\begin{equation}\label{1_52}
kq=- \arctan \frac{4k}{v_0} \Longrightarrow \tan(kq)=-\frac{4k}{v_0}\,.
\end{equation}
This transcendental equation has a countably infinite number of solutions, $k_n$, that give the energy levels corresponding to this situation. 
In the limit $q=0$, we have only one solution, $k=0$. Outside the interval $[0,q]$, we have the equations 
\begin{equation}\label{1_53}
\varphi(0^-)=0 \Longrightarrow A_1+B_1=0;\qquad \varphi'(q^+)= \frac{w_0}{4}\,\varphi(q^+)  \Longrightarrow  \frac{A_3}{B_3}= - e^{2ikq}\,\frac{w_0-4ik}{w_0+4ik}.
\end{equation}
These are relations between coefficients, which do not provide of any further information, so that we shall not 
refer for similar situations which will appear for all other cases.

\medskip

\noindent
\underline{Case 2}: $v_1=1$, $w_1=-1$. 

\begin{figure}[ht]
\centering
\includegraphics[width=0.4\textwidth]{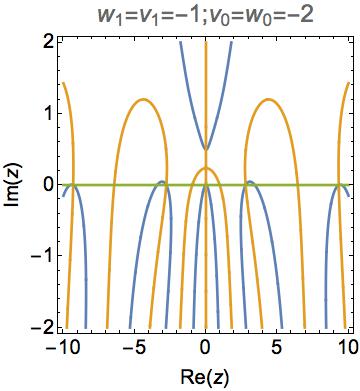}
\quad\qquad
\includegraphics[width=0.4\textwidth]{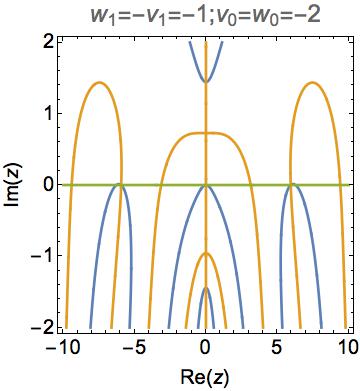}\\
\includegraphics[width=0.15\textwidth]{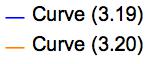}
\caption{\small Plots of the equation ${\rm Re}(\Delta(z))={\rm Im}(\Delta(z))=0$ over the complex $z$-plane when both singularities have couplings (as indicated on the top) in the decoupling regime. The green line represents the axis ${\rm Im}(z)=0$.}
\label{casexc-51}
\end{figure}

\noindent
The boundary conditions are \eqref{1_47} for $x=0$ and \eqref{1_48bis} for $x=q$. The condition at $x=0$ has already been studied, providing equation \eqref{1_49}, so that let us consider the new boundary condition at $x=q$. It comes from $\varphi'(q^-)= -\frac{w_0}{4}\,\varphi(q^-)$:
\begin{equation}\label{1_54}
\frac{A_2}{B_2}= -e^{2ikq}\,\frac{w_0+4ik}{w_0-4ik}= -e^{2ikq} \exp\left( 2i\arctan \frac{4k}{w_0}   \right)\,.
\end{equation}
This equation is to be compared to (\ref{1_49}). The result given in terms of a generalized Lambert equation is:
\begin{equation}\label{1_55}
e^{2ikq}= \frac{v_0-4ik}{v_0+4ik}\,\frac{w_0-4ik}{w_0+4ik}\,.
\end{equation}
This transcendental equation can be written in another way by playing with the formulas for the $\arctan$. Note that from (\ref{1_55}) we obtain straightforwardly the following expression:   
\begin{equation}\label{1_56}
-kq=\arctan \frac{4k}{w_0}+ \arctan \frac{4k}{v_0} \quad \Longrightarrow \quad 
{\tan (kq)= -\frac{4k(w_0+v_0)}{w_0v_0-16 k^2}}\,.
\end{equation}
This new transcendental equation gives another set of energy eigenvalues (see Figure \ref{casexc-51} right).
In the limit $q=0$, we have only one solution: $k=0$. 

\medskip

\noindent 
\underline{Case 3}: $v_1=-1$, $w_1=1$. 

\noindent 
The boundary conditions are  (\ref{1_48}) and  (\ref{1_47bis}). This means $\varphi(0^+)=0$, which implies that $A_2+B_2=0$ or $A_2/B_2=-1$ and $\varphi(q^-)=0$, which is 
\begin{equation}\label{1_58}
A_2e^{-ikq}+B_2e^{ikq}=0 \Longrightarrow \frac{A_2}{B_2}=-e^{2ikq}\,,
\end{equation}
so that
\begin{equation}\label{1_59}
\frac{A_2}{B_2}=-1=-e^{-2ikq} \Longrightarrow {k=\frac{\pi n}{q}}\,.
\end{equation}
Again, in the limit $q=0$, the only solution is $k=0$. 
\medskip

 \noindent
 \underline{Case 4}: $v_1=-1$, $w_1=-1$. 

\noindent 
This correspond to (\ref{1_48}) and (\ref{1_48bis}). The boundary condition at $x=0$ gives ${A_2}/{B_2}=-1$.
The boundary condition at $x=q$ is just that given in (\ref{1_54}). Then, the transcendental equation giving the energy levels is given by
\begin{equation}\label{1_61}
e^{2ikq}=  \frac{w_0-4ik}{w_0+4ik}\,,
\end{equation}
in the form of a generalized Lambert equation, or 
\begin{equation}\label{1_62}
\tan(kq)=-\frac{4k}{w_0}
\end{equation}
in the form of a transcendental equation in terms of the tangent.
This transcendental equation also has a countably infinite number of solutions, $k_n$, that give the energy levels corresponding to this situation. (see Figure \ref{casexc-51} left). 
Once more, in the limit $q=0$,  only the solution $k=0$ remains.

\subsection{Only one $\delta^\prime$ coupling in the decoupling limit}

Let us consider now the four cases in which we have a decoupling and a regular coupling.
\smallskip

\noindent
\underline{Case 1}: $v_1=1,w_1\neq \pm 1$. 

\noindent 
From $v_1=1$, that is, from \eqref{1_47}, we know that the conditions $A_1=-B_1\neq 0$ and \eqref{1_49}
must be satisfied, that is
\begin{equation}\label{4.2.1}
\varphi_1(y)\propto \sin(ky), \qquad
{A_2}= -\frac{v_0-4ik}{v_0+4ik}\ {B_2}.
\end{equation}
Note that the system behaves as if there is an impenetrable barrier at $x=0$. 
The coefficients $A_3,B_3$ are obtained from $A_2,B_2$ as in \eqref{1_10bis}
\begin{equation}\label{1_10tris}
\left( \begin{array}{c} A_3  \\[1ex]  B_3  \end{array}  \right)= 
\left( \begin{array}{cc} 
\displaystyle\frac{2k(1+w_1^2)+iw_0}{2k(1-w_1^2)}  &  \displaystyle e^{2ikq}\,\frac{4kw_1+iw_0}{2k(1-w_1^2)}  \\[2ex]  
\displaystyle  e^{-2ikq}\,\frac{4kw_1-iw_0}{2k(1-w_1^2)}  &  \displaystyle\frac{2k(1+w_1^2)-iw_0}{2k(1-w_1^2)} 
   \end{array}  \right)
\left( \begin{array}{c} A_2  \\[1ex]  B_2  \end{array}  \right)\,.
\end{equation}
All the relevant information about bound states, antibound states and resonances is obtained imposing the so-called {\em purely outgoing boundary condition}, which in our case is $A_3=0$. Using \eqref{4.2.1} and \eqref{1_10tris} we get
\begin{equation}\label{4.2.1new}
e^{2ikq} = \frac{v_0-4ik}{v_0+4ik}\ \frac{2k(1+w_1^2)+iw_0}{4kw_1+iw_0}.
\end{equation}
This equation is the equivalent of $\Delta(k)=0$, where $\Delta(k)$ is given in \eqref{1_32}, which provides the relevant information in the non-decoupling case. Indeed, \eqref{4.2.1new} is obtained making $v_1=1$ in $\Delta(k)=0$. The analysis in this case is similar to the one carried out previously in Section~\ref{sectionboundetc}. 

\begin{figure}[ht]
\centering
\includegraphics[width=0.4\textwidth]{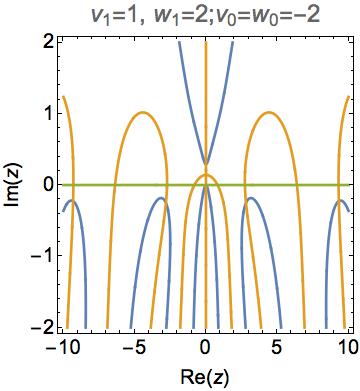}
\quad\qquad
\includegraphics[width=0.4\textwidth]{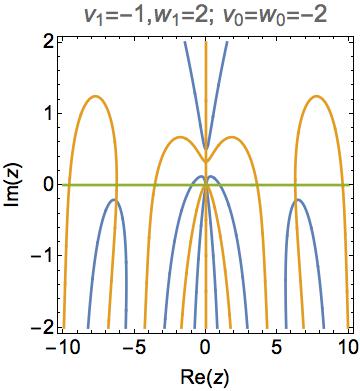}\\
\includegraphics[width=0.15\textwidth]{legend2.jpg}
\caption{\small Plots of the equation ${\rm Re}(\Delta(z)={\rm Im}(\Delta(z)=0$ over the complex $z$-plane when one singularity, as indicated on the top, takes the decoupling value for the coupling. The green line represents the axis ${\rm Im}(z)=0$.}
\label{casexc-52}
\end{figure}

We are  interested  especially in the case where $q\to 0$. Then, from \eqref{4.2.1new} we get
\begin{equation}\label{4.2.1bis}
(v_0+4ik)(4kw_1+iw_0)= (v_0-4ik) (2k(1+w_1^2)+iw_0).
\end{equation}
The complex solutions of this equation are the following:
\begin{equation}\label{4.2.1tris}
k_0=0, \qquad 
k_1=- i \ \frac{4 w_0 + v_0 (1- w_1)^2}{4 (1 + w_1)^2}.
\end{equation}
The solution $k=0$ is not relevant, but the other produces interesting results: If 
$4 w_0 + v_0 (1- w_1)^2<0$, $k_1$ is on the positive imaginary axis and it 
corresponds to a bound state, and if $4 w_0 + v_0 (1- w_1)^2>0$, $k_1$ is on the negative imaginary axis and it corresponds to an anti-bound state. The curves solving the equations ${\rm Re}(\Delta)={\rm Im}(\Delta)=0$ for this case are represented in Figure \ref{casexc-52} left

\medskip

\noindent 
\underline{Case 2}: $v_1=-1$, $w_1\neq \pm 1$. 

\noindent 
The system behaves also as if there were an impenetrable barrier at $x=0$. Now, from $v_1=-1$ (or equivalently from \eqref{1_48}), we find the condition $A_2/B_2=-1$ and consequently \eqref{1_10tris}, which must be also satisfied. Then, the {\em purely outgoing boundary condition} $A_3=0$ implies that (see Figure \ref{casexc-52} right)
\begin{equation}\label{4.case2.1}
e^{2ikq}=\frac{2k(1+w_1^2)+iw_0}{4kw_1+iw_0} .
\end{equation}
In the limit $q\to 0$ we get the condition $k(1-w_1)^2=0$, which has the unique solution $k=0$.

\medskip

\noindent 
\underline{Case 3}: $v_1\neq \pm 1$, $w_1=1$. 

\noindent 
In this situation we must take into account \eqref{1_47bis} and \eqref{1_10}. The first pair of equations shows the presence of a kind of impenetrable barrier, now in $x=q$, and also the fact that $A_2=-e^{2ikq}B_2$. Equation \eqref{1_10} can be rewritten in the form
\begin{equation}\label{4.case3.1}
\left( \begin{array}{c} A_1  \\[1ex]  B_1  \end{array}  \right)= 
K^{-1}M_v^{-1}K
\left( \begin{array}{c} A_2  \\[1ex]  B_2  \end{array}  \right)=
\left( \begin{array}{cc} 
\displaystyle\frac{2k(1+v_1^2)-iv_0}{2k(1-v_1^2)}  &\displaystyle  - \frac{4kv_1+iv_0}{2k(1-v_1^2)}  \\[2ex]  
\displaystyle\frac{-4kv_1+iv_0}{2k(1-v_1^2)}  & \displaystyle \frac{2k(1+v_1^2)+iv_0}{2k(1-v_1^2)} 
   \end{array}  \right)
\left( \begin{array}{c} A_2  \\[1ex]  B_2  \end{array}  \right)\,.
\end{equation}
The {\em purely outgoing boundary condition} is in the present case is $B_1=0$, so that
\begin{equation}\label{4.case3.2}
e^{2ikq}=-\frac{2k(1+v_1^2)+iv_0}{4kv_1-iv_0} .
\end{equation}
In the limit $q\to 0$, we get $k(1+v_1)^2=0$, which as the unique solution $k=0$. Observe that this case is similar to the previous one, Case 2.

\medskip

\noindent 
\underline{Case 4}: $v_1\neq \pm 1$, $w_1=-1$. 

\noindent 
For $w_1=-1$, the second equation in \eqref{1_48bis} imposes $\varphi'(q^-)= -\frac{w_0}{4}\,\varphi(q^-)$. Then
\begin{equation}\label{4.case4.1}
{A_2}= -e^{2ikq}\, \frac{w_0+4ik}{w_0-4 ik}\ {B_2}.
\end{equation}
Taking this into \eqref{4.case3.2}, and imposing the {\em purely outgoing boundary condition} $B_1=0$, we get
\begin{equation}\label{4.case4.2}
e^{2ikq} = -\frac{w_0-4ik}{w_0+4ik}\ \frac{2k(1+v_1^2)+iv_0}{4kv_1-iv_0}.
\end{equation}
In the limit case $q\to 0$, we have
the following complex solutions 
\begin{equation}\label{4.case4.3}
k_0=0, \qquad 
k_1=- i \ \frac{4 v_0 + w_0 (1+v_1)^2}{4 (1 -v_1)^2}.
\end{equation}
As in  Case 1 studied before, the solution $k=0$ is not relevant, but the other one, $k_1$, corresponds either to a bound state if $4 v_0 + w_0 (1+v_1)^2<0$ or to an anti-bound state if $4 v_0 + w_0 (1+v_1)^2>0$.

\section{Concluding remarks}

We have studied a one-dimensional two-point perturbation of the free Hamiltonian $H_0=-d^2/dx^2$ of the form $v_0\delta(x)+2v_1\delta'(x)+w_0\delta(x-q)+2w_1\delta'(x-q)$ with $q>0$. This type of potentials have interest in one dimensional quantum field theory and on the study of the Casimir effect as shown by previous work of our group \cite{JMC,JMG,MM}. We have obtained two remarkable unexpected results.

The former refers to the limit  $q\to 0$. The result is a point potential of the form $u_0\delta(x)+2u_1\delta'(x)$, where $u_0$ and $u_1$ are not the sums $v_0+w_0$ and $v_1+w_1$ respectively, but more complicated functions of these arguments instead. A law of composition for the coefficients of the deltas is produced with structure of group which coincides with the Borel subgroup of $SL_2({\mathbb R})$. From the analysis of reflection and transmission coefficients, we obtain the same group law.

The second one comes from the analysis of the decoupling cases which arises by the choices $v_1=\pm 1$ (decoupling case at $x=0$) and $w_1=\pm 1$ (decoupling case at $x=q$). When considering the decoupling cases at $x=0$ and $x=q$, the barrier seems to be impenetrable at these points so that no scattering states are produced. One finds bound or antibound states as solutions of a generalized Lambert equation on terms of the momentum.     When the decoupling regime is produced at one point only (either $x=0$ or $x=q$), we have impenetrability at this point and scattering through the other one and resonances (as pair of  poles of the analytic continuation of the $S$-matrix in the momentum representation) can be found. However, it is quite remarkable to note that these poles may lie on the real axis if at least one of the $\delta^\prime$ couplings is complex. This situation may violate a widely accepted causality condition \cite{NU}.

Finally, we mention that in the course of this paper we have clarified the relation between two apparently different ways of characterizing self-adjoint extensions of the $H_0$ operator. The first one is based on the matching conditions determined from the Kurasov matrices. We have shown
that they form the Borel subgroup of $SL_2(\mathbb{R})$. The standard approach to deal with self-adjoint extensions of symmetric operators, coming back to von Neumann, is through unitary matrices, see e.g. the previous works \cite{aso-mc1,aso-mc2,mc-kk1} by Asorey, Mu\~noz-Casta\~neda {\it et al}. The connection from these two points of view starts from the isomorphism between the $SL_2(\mathbb{R})$ and $SU_{1,1}(\mathbb{C})$ groups. In fact they are conjugate subgroups inside $GL_2(\mathbb{R})$:
\[
g\, SL_2(\mathbb{R})\, g^{-1}=SU_{1,1}(\mathbb{R}) \quad , \quad g=\left(\begin{array}{cc} 1 & -i \\ 1 & i \end{array}\right) \, .
\] 
The matching conditions (\ref{1_4}) become in the $SU_{1,1}(\mathbb{C})$ framework:
\begin{equation}
\left(\begin{array}{c} \varphi(q^+) - i \varphi^\prime(q^+) \\ \varphi(q^+) + i \varphi^\prime(q^+) \end{array}\right)=W_v \,\left(\begin{array}{c} \varphi(q^-) - i \varphi^\prime(q^-) \\ \varphi(q^-) + i \varphi^\prime(q^-) \end{array}\right), \label{suls} 
\end{equation}
where the $W_v$ is conjugated to the Kurasov matrix $M_v$ through the action of $g$:
\begin{equation}
W_v=g\, M_v\, g^{-1}= \frac{1}{2(1-v_1^2)}\left(\begin{array}{cc} 2(1+v_1^2)-i v_0 & 4v_1-iv_0 \\ 4v_1+iv_0 & 2(1+v_1^2)+i v_0 \end{array} \right)\, .
\end{equation}
In a parallel reshufling of the linear system (\ref{suls}) to that performed to define the $S_q$ scattering matrix from the $T_q$ matrix
and passing from the (\ref{1_13}) equation to (\ref{1_24}) and (\ref{1_251}) we rewrite (\ref{suls}) in the form
\begin{equation}
\left(\begin{array}{c} \varphi(q^-) - i \varphi^\prime(q^-) \\ \varphi(q^+) + i \varphi^\prime(q^+) \end{array}\right)=U_v \, \left(\begin{array}{c} \varphi(q^+) - i \varphi^\prime(q^+) \\ \varphi(q^-) + i \varphi^\prime(q^-) \end{array}\right), \label{uls}
\end{equation}
where the unitary matrix $U_v$ is obtained from $W_v$ and reads
\begin{equation}
U_v=\frac{1}{W_v^{11}}\left(\begin{array}{cc} 1 & - W_v^{12}  \\ W_v^{21} & {\rm det}W_v\end{array}\right)=\frac{2(1-v_1^2)}{2(1+v_1^2)-i v_0}\left(\begin{array}{cc} 1 & \frac{-4v_1+iv_0}{2(1-v_1^2)} \\ \frac{4v_1+iv_0}{2(1-v_1^2)} & 1 \end{array}\right) \label{usa} \, .
\end{equation}
In this indirect way the Kurasov matrices determining the matching conditions that define the $\delta$-$\delta^\prime$ interactions are related 
to a subset of the $U(2)$ group which in turn characterizes a variety of self-adjoint extensions of the $H_0$ operator in the standard manner.

A remarkable fact is the following: even though $M_v$ and $W_v$ are singular matrices at the decoupling limit $v_1=\pm 1$ the corresponding 
unitary matrices are regular:
\begin{equation}\label{usl}
U_v \Big\vert_{v_1=1}= \left(\begin{array}{cc} 0 & -1 \\ \frac{4+iv_0}{4-iv_0} & 0\end{array}\right) \ , \qquad U_v \Big\vert_{v_1=-1}= \left(\begin{array}{cc} 0 & \frac{4+iv_0}{4-iv_0} \\ -1 & 0\end{array}\right) \, .
\end{equation} 
Writing the linear system (\ref{usa}) for the decoupling limit of the coupling $v_1=1$ we obtain two equations. First,
\begin{equation}
-(\varphi(q^-)+i \varphi^\prime(q^-))=\varphi(q^-)-i \varphi^\prime(q^-) \, \, \Rightarrow \, \, \varphi(q^-)=0 \, \, ,
\end{equation}
i.e., Dirichlet boundary conditions are satisfied reaching the point $q$ from the left. Second,
\begin{equation}
\frac{4+iv_0}{4-i v_0}(\varphi(q^+)-i \varphi^\prime(q^+))=\varphi(q^+)+i \varphi^\prime(q^+) \, \, \Rightarrow \, \, \varphi(q^+)- \frac{4}{v_0}\varphi^\prime(q^+)=0 \, \, 
\end{equation}
sets Robin boundary conditions at $q$ coming from the right. If the other decoupling value for the coupling, $v_1=-1$, is chosen the situation is identical but the conditions at $q$ from the left or from the right are exchanged. The results above is in perfect agreement with the Kurasov matching conditions at decoupling limit of the couplings, as expressed in formula \eqref{1_47}. Finally, we mention that in the usual treatment this situation is determined from diagonal rather than anti-diagonal matrices. To meet this criterion one merely multiply (\ref{usl}) by the $\sigma_1$ Pauli matrix.

Finally, we extract some physical consequences from the non-abelian superposition law for the $\delta$-$\delta^\prime$ potentials:
\begin{enumerate}
\item The quantum self-energy  of two $\delta$-$\delta^\prime$ configurations due to quantum vacuum scalar fluctuations should inherit somehow some features from the non-abelian composition law. A wel known procedure to regularize such divergent quantity is to start from the heat trace of the Hamiltonian operator: $h_H(t)={\rm Tr}_{L^2}{\rm exp}[-t H]$, where $t$ is Schwinger proper time. This spectral function is obtained for any Schr$\ddot{\rm o}$dinger Hamiltonian in terms of the bound state energies and the spectral density, defined in turn from the total phase shift, see the two formulas just above (32) in \cite{JMG}. From the total phase shift produced by the $\delta$-$\delta^\prime$ potential determined by the $M_v$ Kurasov matrix and the bound state energy we find:
\begin{equation}
h_H\left(t;\,M_v\right)=e^{\frac{t  v_0^2}{4 \left(v_1^2+1\right){}^2}}
   \left(2 i \pi  \,\text{erfc}\left(\frac{\sqrt{t
   } v_0}{2 \left(v_1^2+1\right)}\right)+\theta
   \left(-v_0\right)-4 i \pi \right),
\end{equation}
where ${\rm erfc}$ is the complementary error function and $\theta$ the Heaviside step function. It is of note that the exact heat kernel for a single $\delta$-$\delta^\prime$ is well defined in the decoupling limit $v_1\rightarrow\pm 1$. We conjecture that taking the zero distance limit in the double $\delta$-$\delta^\prime$ potential, the heat trace for the superposed potential will be of the form $h_H\left(t;\,M_v\cdot M_w\right)=h_H\left(t;\,M_u\right)$. The quantum vacuum energy is essentially the spectral zeta function evaluated at $s=-1/2$. This second spectral function is obtained from the heat trace via Mellin's transform of the heat trace: $\zeta_H(s)=\frac{1}{\Gamma(s)}\int_0^\infty dt \,\,t^{s-1}h_H(t)$,  where $\Gamma(s)$ is Euler Gamma function and $s$ a complex parameter.
$s=-1/2$ is a pole of the meromorphic function $\zeta_H(s)$ in $\mathbb{C}$. One regularizes the divergent quantum vacuum energy by assigning to $\zeta_H$ its value in a regular point. What the non-abelian addition law tell us are the values of the parameters $u_0$ and $u_1$ as functions of $v_0,w_0,v_1,w_1$ entering in this regularized expression after the $q=0$ limit has been taken. 

\item An ideal model of electric conductivity in solids is provided by a $\delta$-Dirac comb  where $\delta$-point interactions sit at  the ions sites. This model can be enriched in a twofold way: (1) $\delta^\prime$ potentials are added at every site of the lattice. (2) Two species of ions, henceforth two species of $\delta$-$\delta^\prime$ interactions, such that the solid is characterized by the periodic potential 
\begin{equation}
V(x)=\sum_{n\in\mathbb{Z}}\left(v_0\delta(x-n q)+v_1\delta(x-nq)+u_0\delta(x-n q+p)+u_1\delta(x-nq+p)\right) \label{gdiraccomb}
\end{equation}
are considered. There are two limits to a single species: $p\rightarrow 0,q$. Unlike in the Dirac comb of two species where the two limits are equivalent, the two merging processes are different in the (\ref{gdiraccomb}) comb due to the non-abelian superposition law. In fact,
allowing $p$ to vary we have an infinitely repeated Casimir piston, which can be reduced to one piston by restricting the system to the primitive cell.
\end{enumerate}

\section*{Acknowledgements}

We acknowledge the financial support of the Spanish MINECO (Project MTM2014-57129-C2-1-P) and Junta de Castilla y
Le\'on (UIC 011). JMMC would like to acknowledge the fruitful discussions with Michael Bordag, Klaus Kirsten and Manuel Asorey.

\bibliographystyle{unsrt}

\bibliography{addition-ddp}

\end{document}